\documentclass{article}
\usepackage[utf8]{inputenc}
\usepackage[T1]{fontenc}
\usepackage[english]{babel}
\usepackage{graphicx}
\usepackage{floatflt}
\usepackage{amsthm}
\usepackage{amsmath}
\usepackage{amssymb} 
\usepackage[usenames, dvipsnames]{color}
\usepackage{mathrsfs}
\usepackage{geometry}  
\usepackage{makeidx}
\usepackage{physics}
\usepackage{mathtools}
\usepackage{authblk}
\usepackage{lineno}
\usepackage[numbers,sort&compress]{natbib}
\usepackage{csquotes}
\usepackage{chngcntr}
\usepackage{setspace}
\usepackage[labelfont=bf]{caption}
\usepackage{comment}
\usepackage{ulem}

\DeclareUnicodeCharacter{03BC}{$\mu$}
\DeclareUnicodeCharacter{2009}{ }
\DeclareUnicodeCharacter{03BB}{\ensuremath{\lambda}}
\DeclareUnicodeCharacter{223C}{\ensuremath{\sim}}

\title{Quantum Walk Comb in a Dual Waveguide Quantum Cascade Laser}

\author[1,*]{Alessio Cargioli}
\author[1]{Miguel Montesinos Ballester}
\author[2]{Sonja Gantner}
\author[2]{Emilio Gini}
\author[1]{Mattias Beck}
\author[1]{Jerome Faist}

\affil[1]{Institute for Quantum Electronics, ETH Zurich, CH-8093 Zurich, Switzerland}
\affil[2]{FIRST - Center for Micro- and Nanoscience, ETH Zurich, CH-8093 Zurich, Switzerland}

\affil[*]{Contact Email: acargioli@phys.ethz.ch}

\date{}

\modulolinenumbers[1]

\begin{document}
\renewcommand{\figurename}{Figure}
\def\equationautorefname#1#2\null{Eq.#1(#2\null)}
\maketitle
\numberwithin{equation}{section}

\begin{abstract}
Ring quantum cascade lasers (QCLs) proved to be a versatile tool for generating tunable and stable frequency combs in the mid infrared range in the form of quantum walk combs. By homogeneously integrating a racetrack QCL with a passive waveguide, which lays on top of the active region plane and therefore can be designed to be fully independent from the laser geometry, we improve the light outcoupling from the ring by more than 2 orders of magnitude reaching a maximum output power of 120 mW. In addition, we show that it is possible to achieve quantum walk comb operation in the devices under analysis. Finally, we prove that we can change the light dispersion by tuning the parameters of the passive waveguide, with a direct impact on the behavior of the generated comb.

\end{abstract}

\setcounter{section}{1}
\counterwithout{equation}{section}

\newpage

\section*{Introduction}

Frequency combs have become fundamental tools for many research fields \cite{fortier_20_2019} from spectroscopy \cite{picque_frequency_2019}, to telecommunications \cite{hussein_passive_2024, okawachi_chip-scale_2023} and high precision metrology \cite{udem_optical_2002, ye_optical_2003}. Therefore, it is of great importance to obtain compact and bright frequency comb sources. To this purpose, Quantum Cascade Lasers (QCLs) \cite{faist_quantum_1994} offer an interesting solution in the Mid-Infrared (Mid-IR) and THz spectral range. Since the first proof of self-starting combs in Fabry-Perot QCLs \cite{hugi_mid-infrared_2012, burghoff_terahertz_2014}, the field expanded both from a fundamental understanding of the physics of the system \cite{burghoff_unraveling_2020, opacak_theory_2019}, which consists of a complex interplay of dispersion, Kerr nonlinearity, and an ultra-fast gain recovery time, and from a technological point of view, with the aim of achieving broader and more stable combs \cite{villares_dispersion_2016, kapsalidis_mid-infrared_2021}. More recently, the study of QCL ring cavities has attracted increasing attention, not only because these platforms can show self-starting frequency comb formation as well \cite{meng_mid-infrared_2020, jaidl_comb_2021}, but because the inherent absence of spatial hole burning allows the study of a different physics inside the laser \cite{lugiato_nonlinear_2015}, opposed to a classical Fabry-Perot geometry. It was shown that it is possible to obtain self-starting solitons in both Mid-IR \cite{meng_dissipative_2022, kazakov_active_2024, opacak_nozakibekki_2024} and THz \cite{micheletti_terahertz_2023} spectral region. One of the requirements to reach this condition appears to be related to the amount of backscattering of the cavity \cite{seitner_backscattering-induced_2024}, which is typically non negligible. Indeed, if one manages to produce a clean ring with very low backscattering, the typical state of the laser will instead be a monochromatic emission. This result was the starting point for proving the existence of the first Quantum Walk Comb (QWC) in the 8$\mu$m spectral region \cite{heckelmann_quantum_2023}. By using a Radio-Frequency (RF) current modulation at the roundtrip frequency of the cavity, it is possible to generate a tunable comb, fully controllable by the power and by the detuning of the modulation. This new laser state can be accurately described by a complex Ginzburg-Landau equation \cite{dikopoltsev_theory_nodate} and it is characteristic of systems as QCLs due to their fast gain recovery time which tends to suppress amplitude fluctuations, favoring a frequency-modulated frequency comb formation. Despite these extremely impactful results, the applications of QWCs remain limited mainly because of an inherently low output power due to the necessity of a ring geometry. It was recently shown that, by coupling a QCL racetrack (RC) with an active bus element that could be biased independently, it was possible to reach a QWC operation with an extraction power up to 100 mW \cite{letsou_high-power_2025}. This is an important step towards high power and highly controllable on chip frequency combs but presents the limitation of power dissipation. Indeed, if the aim is to achieve a Mid-IR photonic integrated circuit (PIC), there is the necessity of shifting to a coupling to passive elements. Up to this day, Mid-IR photonic integration remains very difficult and it is mostly achieved through heterogeneous integration on silicon \cite{lin_mid-infrared_2018,spott_quantum_2016, pelucchi_potential_2022} with limitations which are mainly posed by the thermal management and by the complicated alignment of the laser to the passive elements. It is more appealing to look for an homogeneous integration where it is possible to integrate passive and active elements during the same fabrication process. Some works showed how this can be done by coupling a Fabry-Perot QCL to a plasmonic waveguide \cite{hinkov_mid-infrared_2022}, or through a but-coupled  InGaAs waveguide \cite{wang_monolithic_2022}, or by evanescence coupling with an InGaAs waveguide grown below the QCL active material \cite{jung_homogeneous_2019, burghart_multi-color_2025}. Using InGaAs as passive material remains a promising idea due to the relatively low losses in the Mid-IR region \cite{montesinos-ballester_low-loss_2024} as opposed to a plasmonic element which is better suited as a sensing area. In this work we couple a QCL racetrack to a passive InGaAs waveguide, which is regrown on top of the active material via Metal Organic Vapor Phase Epitaxy (MOVPE), in order to extract the light trapped inside the ring, reaching a maximum output power of 120 mW. We prove it is possible to achieve a clean quantum walk comb state and that we can strongly influence the comb formation by acting on the dispersion by changing the width of the passive waveguide \cite{bidaux_coupled-waveguides_2018}. The ability of redirecting the light generated in the ring through a structure which is totally independent of the geometry  of the active region opens up the possibility of monolithically integrating arbitrary passive elements on the same laser chip, pointing towards a plethora of applications that can benefit from having one, or more, fully tunable quantum walk combs on the same chip. 

\section*{Results and Discussion}

\begin{figure}[htb!]
  \centering
  \includegraphics[width=\textwidth]{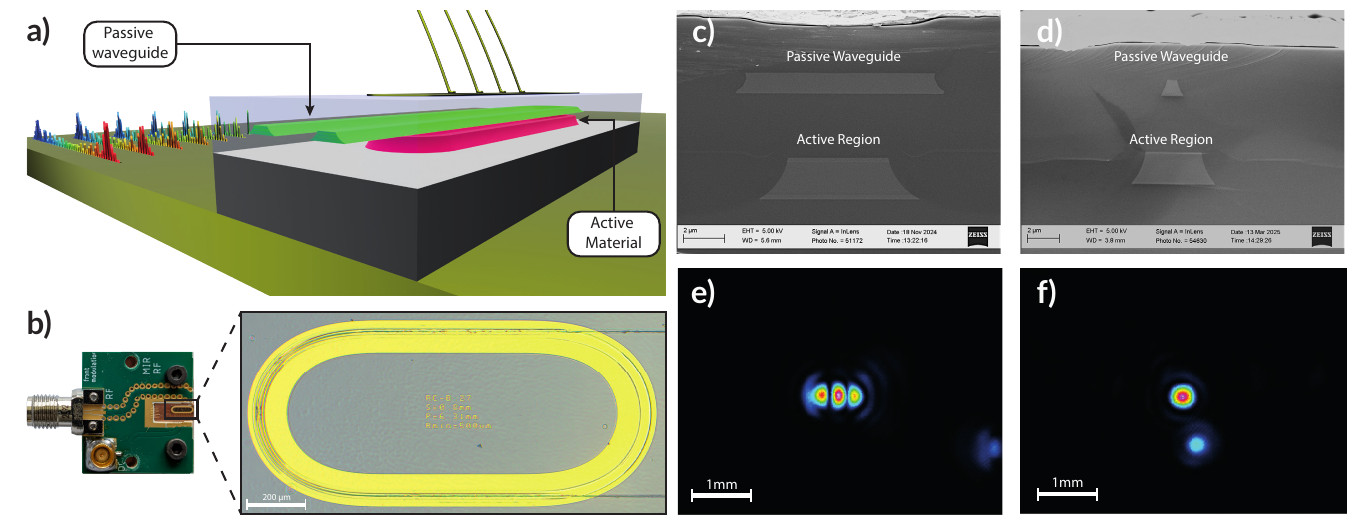}
  \caption{\textbf{a)} 3D schematic of the double waveguide laser. \textbf{b)} Picture of a double waveguide laser mounted on a a standard PCB, and respective top-view showing the electrical contact and the trace of the U-shaped passive waveguide. \textbf{c-d)} SEM pictures of a cross section of a double waveguide laser with different passive waveguide widths. \textbf{e-f)} Far field measurements of the emitted light at the facet from the passive waveguide, wider and narrower than the active region, respectively. }
  \label{fig:render}
\end{figure}

The laser fabrication consists of a variation of the buried heterostructure process \cite{beck_continuous_2002} which extensively uses MOVPE regrowths to define the different layers of the device. The active region is first patterned by wet-etching, and subsequently, an insulating layer of Fe-doped InP is regrown laterally. Afterwards, a layer of conductive Si-doped InP and Si-doped InGaAs is grown and then patterned into a shape that can be chosen independently from the active material geometry. Finally, a layer of Si-doped InP is grown to complete the burying of the structure. We refer to the supplementary material for a more detailed description of the processing steps (S\ref{SI:device}). A schematic of the final device, with an U-shaped passive waveguide as an example, is reproduced in the 3D model in Fig.\ref{fig:render}.a . 

The final device mounted on a printed circuit board (PCB), which is needed to independently bias the laser through a DC and RF line, is shown in Fig.\ref{fig:render}.b where a top view of a finished device is also displayed. Below the gold contact, it is possible to distinguish the racetrack shape, corresponding to the active region, and the U-shaped passive waveguide which allows extraction of the light at the facet. We report two SEM images of the cross section of a double waveguide laser cleaved in its central region in Fig.\ref{fig:render}.c-d, where we show that, even if the laser has undergone multiple regrowth, it is possible to achieve a flat InGaAs structure that can be well aligned to the bottom active region, and that it is possible to tune both the width of the passive element, and the spacing of the two waveguides which can be freely chosen by the MOVPE regrowth. Depending on the top waveguide size, we can obtain different far fields. In particular, with reference to the active region, if a wider top waveguide (for brevity it will be referred to as wide waveguide) is chosen, the far field will have the shape of a higher-order mode of the passive element (Fig.\ref{fig:render}.e). This is due to the laser wavelength, which in our case is around 4.5 $\mu$m, smaller than the waveguide width which is around 8 $\mu$m. Instead, if a narrower top waveguide (for brevity it will be referred to as narrow waveguide) is chosen, a fundamental TM$_{00}$ mode is reproduced. Clearly, guiding the light via a clear path allows better control of the beam shape compared to the simpler case of a single ring, where the light is only extracted via bending losses\cite{heckelmann_quantum_2023}.

\begin{figure}[htb!]
  \centering
  \includegraphics[width=\textwidth]{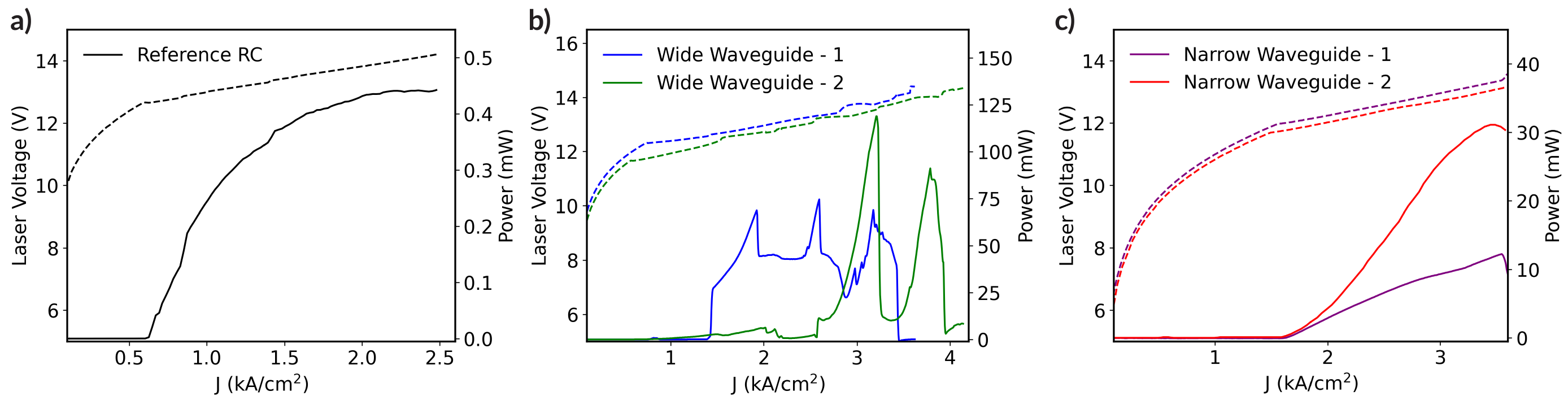}
  \caption{Laser voltage vs current density (dashed line) and laser power vs current density (solid line) for: \textbf{a)} one reference racetrack without passive waveguide, \textbf{b)} two racetracks with wider top waveguide then the active region, \textbf{c)} two racetracks with a narrower top waveguide than the active region. }
  \label{fig:liv}
\end{figure}

The size of the waveguide has an impact not only on the mode shape but also on the total outcoupled power. We report the light-current density and voltage-current density characteristics for a reference RC without any top waveguide (Fig.\ref{fig:liv}.a), for the case of a wide waveguide (Fig.\ref{fig:liv}.b) and for a narrow waveguide (Fig.\ref{fig:liv}.c). All measurements are performed with a DC current source and at a heat-sink temperature of 253 °K.

\begin{figure}[htb!]
  \centering
  \includegraphics[width=\textwidth]{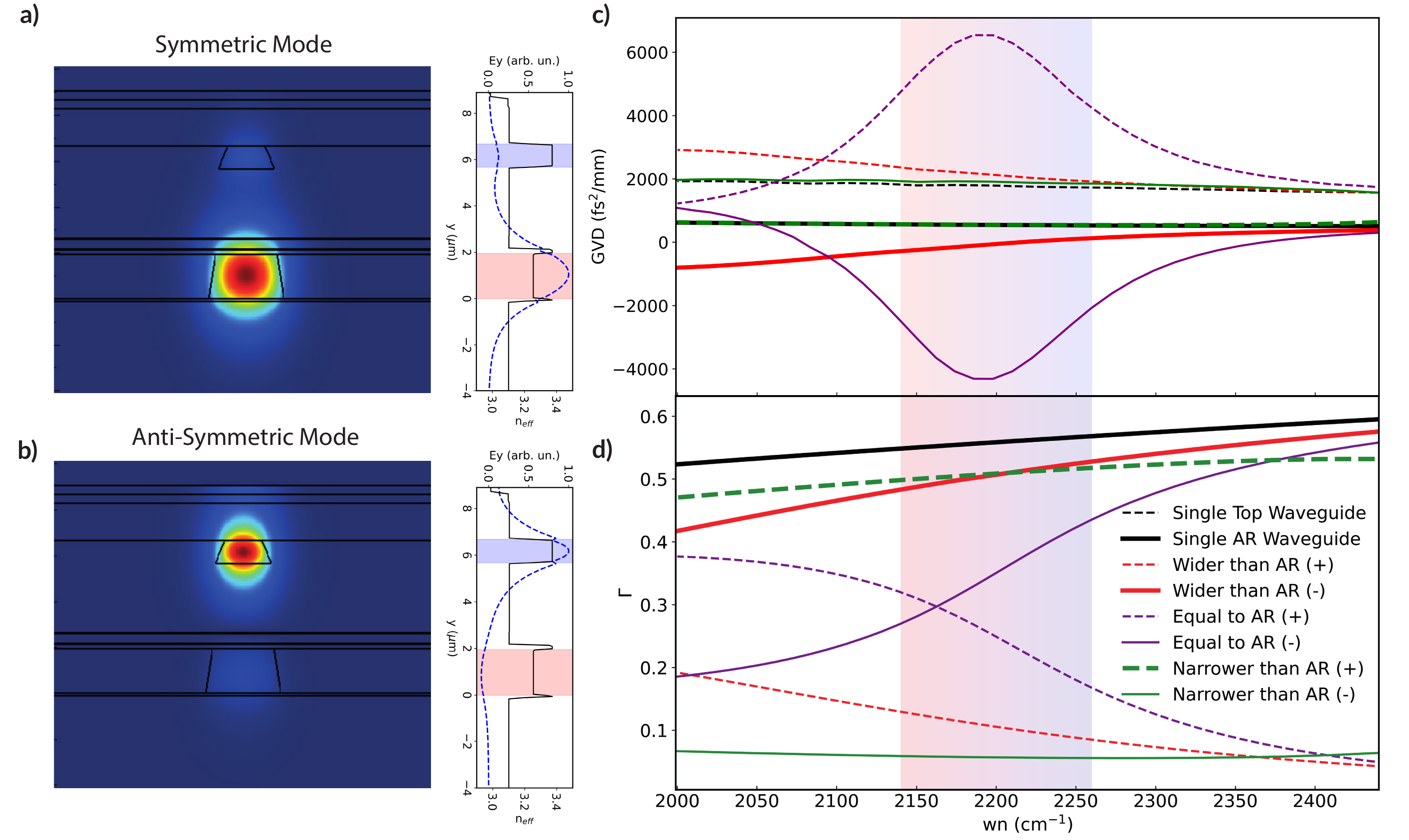}
  \caption{\textbf{a-b)} Symmetric and Anti-Symmetric TM mode solution, respectively, for a coupled waveguide system. Each inset, represents the mode profile in the center of the structure and the relative refractive index profile. The red-shaded and purple-shaded region, correspond to the active region and passive waveguide position, respectively. \textbf{c-d)} Simulated Group Velocity Dispersion (GVD) and mode overlap with the active material ($\Gamma$), for the two uncoupled waveguides and for the cases of a wider, narrower and same size top waveguide compared to the active region width. Both symmetric (+) and antisymmetric (-) solutions are displayed. The dispersion and the overlap factor of the lasing modes are indicated by thicker lines. The shaded region corresponds to the gain bandwidth of the active material under consideration. For all curves the active region width is chosen to be 5 $\mu$m. The top waveguide width is 3 $\mu$m and 8.5 $\mu$m for the narrow and wide waveguide, respectively.}
  \label{fig:dispersion}
\end{figure}

A typical output power for a single ring is on the order of hundreds of $\mu$W while the lasers with the extraction mechanism present typical values of tens of mW, up to a maximum of 120 mW, showing an increase in extraction efficiency of more than 2 orders of magnitude. In addition, if we compare the voltage-current curves for the reference RC and the wide waveguide laser, which are processed on a similar active material (see Supp. Mat. S\ref{SI:LIV}), we can see that there is no significant increase in the current threshold, indicating that the material regrowth is optimal, it does not impact the heat dissipation, and it does not introduce any additional waveguide losses. It is clear from the comparison of the light-current curve between the wide waveguide and the narrow waveguide devices, that the first ones present a higher irregularity of the emission power. We believe these high power oscillations to be caused by the mode mismatch happening at the interface between the region where the two cavity modes are coupled and the region where only the passive element is present. This also causes the far field of the wide waveguide lasers to be a higher order cavity mode even if the supermode shape is insensitive to the top waveguide width in the length range under investigation and it is as the one reported in Fig.\ref{fig:dispersion}.a-b and in S\ref{SI:Sim1}, presenting a single maximum in the InGaAs region. This explanation is corroborated by the very regular power-current curves of the narrow waveguide lasers which present a fundamental TM$_{00}$ mode as far field.

Changing the width of the top waveguide greatly affects not only the mode shape and the outcoupling efficiency, but also the light dispersion. In Fig.\ref{fig:dispersion}.c-d we report the simulated group velocity dispersion (GVD) and the active region overlap factor ($\Gamma$) for the relevant cases. In particular, considering the gain bandwidth of the active material under consideration, which roughly spans from 2140 cm$^{-1}$ to 2260 cm$^{-1}$, it is possible to pass from the narrow waveguide case, where the GVD of the symmetric and antisymmetric modes result to be very similar to the dispersion of the uncoupled cavities, to the case where the two widths are equal, where the dispersion presents a resonant behavior at the frequency of interest, and finally to the wide waveguide condition where the resonance is pushed at lower wavenumbers, and in the region of interest we obtain a GVD that crosses 0 but with a significant slope. Obviously, while tailoring the dispersion, it is important to keep under control the overlap factor which, ideally, should be as close as possible to the one of the uncoupled cavity. It is clear from Fig.\ref{fig:dispersion}.d that the only reasonable waveguide widths are those that are far away enough from the completely resonant case, where the two waveguide are equal. For this reason, the two typologies of devices under consideration are the narrow and wide waveguide laser. We underline that, for the narrow waveguide, the lasing mode is the symmetric one, while for the wide waveguide is the antisymmetric one. The change clearly occurs at the resonant width, where there is a frequency at which the favorable overlap factor goes from the symmetric to the antisymmetric supermode. 

\begin{figure}[htb!]
  \centering
  \includegraphics[width=\textwidth]{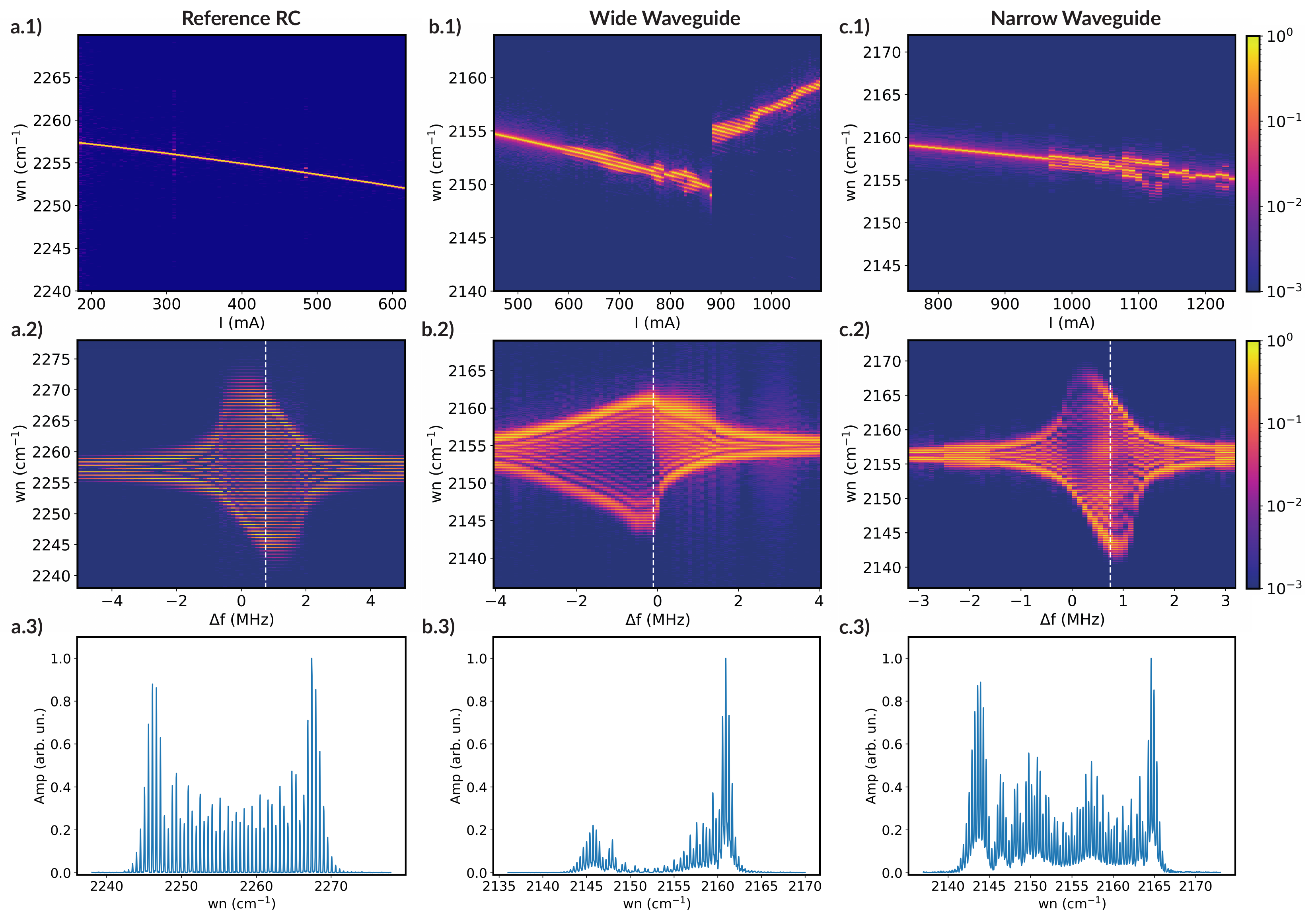}
  \caption{Spectra of a Reference RC \textbf{a)}, of a wide waveguide laser \textbf{b)} and of a narrow waveguide laser \textbf{c)} as a function of the DC current (first row), and as a function of the RF frequency (second row). For each frequency sweep, the broadest spectrum is displayed (third row) corresponding at the position indicated by the vertical white line. The details regarding the RF modulation condition are reported in Methods.}
  \label{fig:spectra}
\end{figure}

Finally, we report a comparison of the spectral behavior between the reference RC, the wide and narrow waveguide lasers. As previously discussed, the reference racetrack spectrum as a function of the bias current (Fig.\ref{fig:spectra}.a.1) is indeed monochromatic, as expected for a clean ring cavity. Regarding the RF modulation response, a spectral map as a function of the modulation frequency, which is swept across the repetition frequency of the laser, is reported in Fig.\ref{fig:spectra}.a.2. The tuning shape is the one expected as reported in \cite{heckelmann_quantum_2023} with a maximum bandwidth of about 30 cm$^{-1}$ as shown in the spectral slice in Fig.\ref{fig:spectra}.a.3.

The CW spectral map of the wide waveguide laser results in a less clean tuning Fig.(\ref{fig:spectra}.b.1), presenting multiple regions where the spectrum is multimode. We attribute this behavior to the non-negligible back-reflections in the counterpropagating direction caused by the finite reflectivity at the extraction facet. This effect was mitigated by evaporating a $\lambda$/4 film of Al$_2$O$_3$ to lower the facet reflectivity from 27$\%$ to roughly 8$\%$. The coating was used on both double waveguide devices under consideration, and we refer to the Supplementary Material for further discussion on the effect of back-reflections (S\ref{SI:spectra}). In general, it is still possible to obtain an operating range where the emission is monochromatic, and therefore we are able to test the response of the device under modulation as reported in Fig.\ref{fig:spectra}.b.2. The expansion resembles a standard one, until the resonant frequency is reached, where a collapse of the spectrum is present. This behavior is caused by a high third order dispersion term \cite{heckelmann_quantum_2023, dikopoltsev_theory_nodate, opacak_impact_2024} which, as previously discussed, is more pronounced for this device if compared to the typical GVD slope of an uncoupled waveguide. Lastly, concerning the narrow waveguide laser, the CW spectral behavior is more regular, indicating that a smaller waveguide is perturbing less the steady state of the laser, even if the effect of the back-reflections is still present at higher currents. In addition, the frequency spectral dependency better resembles the one of the reference device, as we would expect, since the dispersion was chosen to be similar to the one of the uncoupled cavity. 

\section*{Conclusions}
By homogeneously integrating a passive InGaAs waveguide with a QCL racetrack geometry, we proved that it is possible to improve the light extraction efficiency by two orders of magnitude compared to the single ring case. In addition, a quantum walk comb expansion was proven possible for both double waveguide geometries under consideration. By tuning the width of the passive element, it is also possible to change the light dispersion, highly impacting the comb shape, which is influenced by higher order dispersion terms. In general, to the purpose of reaching a reliable and tunable comb condition, the narrow waveguide devices are the better suited ones. The narrower cavity impacts less the dispersion, and presents a better mode matching between the coupled cavity region and the passive element one, as proven by the far field measurement and by the regularity of the Power-Current curve. The back-reflections induced by the facet reflectivity remain problematic for the stability of the CW state of the laser, which can be mitigated by an appropriate anti-reflection coating. With the aim of achieving a fully homogeneous active-passive integration, the back-reflection issue would then become less relevant, since there would not be facets with high reflectivity. In particular, this work can be a first step towards a monolithic active-passive integration with possible applications in spectroscopy by using the advantages posed by the tunability of the quantum walk combs. In addition, the ability to design completely independent passive structures can be exploited to integrate photonic components that are nowadays common in the telecommunication spectral range \cite{elshaari_hybrid_2020, pelucchi_potential_2022}, and particularly appealing would be applications for on chip dispersion compensation for generating short pulses\cite{taschler_short_2023}.

\section*{Methods}

\subsection*{LIV Measurements}
The light from the QCL is collimated using a high numerical aperture lens (Thorlabs C037TME-E) and directed onto a powermeter (Ophir Starlite). The DC current is supplied via QCL2000 (Wavelength Electronics).

\subsection*{Spectral Map}
The spectral maps of the reference racetrack are measured with a commercial fourier transform interferometer (FTIR) Bruker-VertexV80 using an external nitrogen cooled HgCdTe detector. The double waveguide devices, instead, are measured with a home-made FTIR. 

\subsection*{RF Modulation}
The PCB allows an independent biasing of the laser via two ports. One is used for biasing the laser with a DC current, while the other one, is optimized for RF modulation. The RF signal is generated with a Rohde$\&$Schwarz (SMF100A), sent to an RF amplifier (Mini-Circuits ZVE-3W-183+). Because the RF port is shorted with the DC port through the laser, we use a DC-block to protect the Amplifier. The wide waveguide laser characteristic is reported in the supplementary material (S\ref{SI:LIV}). The modulation conditions are : I = 540 mA, $f_{rep} = 11.091$ GHz, $P = 22$ dBm. The narrow waveguide laser spectra is the one of the narrow waveguide-2 reported in Fig.\ref{fig:liv}.c. The modulation conditions are : I = 666 mA, $f_{rep} = 10.15$ GHz., $P = 23$ dBm. The reference RC modulation conditions are: I = 1014 mA, $f_{rep} = 15.691$ GHz., $P = 27$ dBm.
 
\newpage

\section*{Acknowledgements}
The authors gratefully acknowledge founding from the ClosedLoop-LM project (SFA Advanced Manufactoring 2021-2024/ETH), from the MIRAQLS project (European Project 101070700), from the Swiss National Science Foundation (Probing and engineering cavity-mediated vacuum field interactions, 200021-227521)

\section*{Author contributions}
A. Cargioli designed, fabricated, tested the devices and wrote the manuscript. M. M. Montesinos helped in the devices simulation. S. Gantner and E. Gini took care of the MOVPE regrowths. M. Back provided the active material. J. Faist conceptualized and supervised the project. All authors contributed to the review of the manuscript.

\section*{Competing interests}
The authors declare no competing interest.

\section*{Supplementary Information}
Supplementary information is available 

\section*{Data Availability}
The data of this work is available upon reasonable request.

\newpage
\section*{Quantum Walk Comb in a Dual Waveguide Quantum Cascade Laser (Supplementary Material)}
\renewcommand{\figurename}{S}

\counterwithout{figure}{section}
\counterwithout{equation}{section}
\setcounter{figure}{0}

\section*{Detailed Device Fabrication}

\begin{figure}[htb!]
  \centering
  \includegraphics[width=\textwidth]{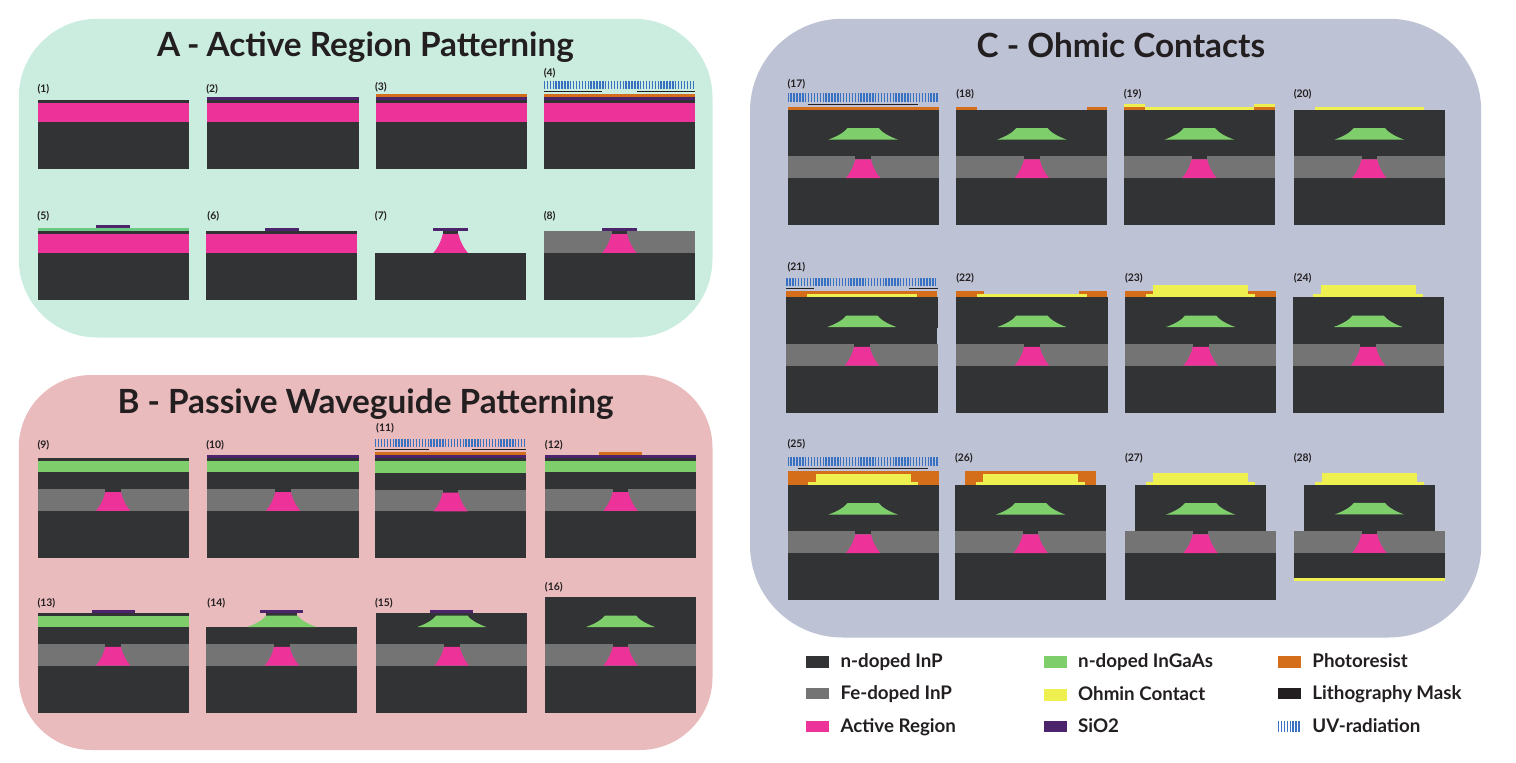}
  \caption{Detailed processing steps of the dual waveguide laser. }
  \label{SI:device}
\end{figure}

We report a detailed description of the processing steps used to produce a dual waveguide laser. In the description, we will refer to the insets of S.\ref{SI:device}. 
The process is divided into 3 main parts. The first one (A) consists of the patterning of the active region (AR), the second one (B) of the definition of the passive waveguide geometry, and finally the third one (C) of the definition of the electrical contacts.

Once the active region is grown via Molecular Beam Epitaxy (MBE), a 500 nm layer of Si-doped InP is regrown on top of it to allow a better subsequent MOVPE regrowth (A.1). The first processing step consists of an deposition of 600 nm of SiO$_2$ which will act as hard mask (A.2). Via contact lithography the pattern of the active region is defined (A.2-5) and the hard mask is patterned (A.6) via reactive ion etching (RIE). After stripping the resist, the active material is then wet-etched using an HBr:Br:H$_2$O solution (A.7). Finally, a Fe-doped InP layer (target doping 6e16/cm$^3$) is laterally regrown (A.8) to achieve electrical insulation.
The hard mask is then stripped, and a layer consisting of Si-doped InP/InGaAs/InP is regrown on top (B.9). Following the same procedure of section A, a SiO$_2$ hard mask is deposited on the surface (B.10), and by contact lithography the passive waveguide structures are defined (B.11-12). The hard mask is again etched via RIE (B.13) and, after resist removal, the structures are wet-etched in the same was as for the AR (B.14). Finally, a Si-doped InP regrowth is done to cover the waveguide laterally (B.15), and then, after hard mask removal, a final layer of Si-doped InP is regrowth on the top (B.16), concluding the burring of the structure.  
The contact definition consists of a standard liftoff of an evaporated layer of Ti/Pt/Au (C.17-20), then a 4 $\mu$m thick layer of gold is deposited on top of the contacts via electroplating to allow a better heat dissipation (C21-24). Finally, an HCl etch is performed to separate the different devices on the chip, the substrate is thinned, and a back-contact of Ge/Au/Ni/Au is evaporated (C.25-28).  
We report a table containing the thicknesses and the doping for wider top waveguide (WTW) and narrower top waveguide (NTW) devices.

\begin{table}[htb!]
    \centering
    \begin{tabular}{l|c|c}
         Material & Thickness ($\mu$m) & Si-Doping (cm$^{-3}$) \\
         \hline
         Au             &   $\sim 4$       &           \\
         InP            &   0.4     &   3e18    \\
         InP            &   0.4     &   2e17    \\
         InP            &   1.1     &   1e16    \\
         InP-Plan       &   0.5     &   1e16    \\
         InGaAs-Wvg     &   1       &   1e16    \\
         InP-Buffer     &   3/2.6   &   1e16    \\
         InP-Plan       &   0.5     &   1e17    \\
         AR             &   1.98    &   -       \\
         InP-Substrate  &   $\sim 220$ &  2e17    \\
         
    \end{tabular}
    \caption{For every layer, a description of the material, the thickness and the doping are reported. The Planarization layer (Plan) is needed in to allow a better MOVPE regrowth after etch. The thickness of the InGaAs waveguide (Wvg) is in both wide waveguide and narrow waveguide the same, while the spacing between the AR and the passive element is slightly tuned, in particular 3 $\mu$m  is chosen for the narrow waveguide, while 2.6 $\mu$m is for the wide waveguide.}
    \label{tab:dw}
\end{table}

\section*{Light-Current-Power measurements}
\begin{figure}[htb!]
  \centering
  \includegraphics[width=.9\textwidth]{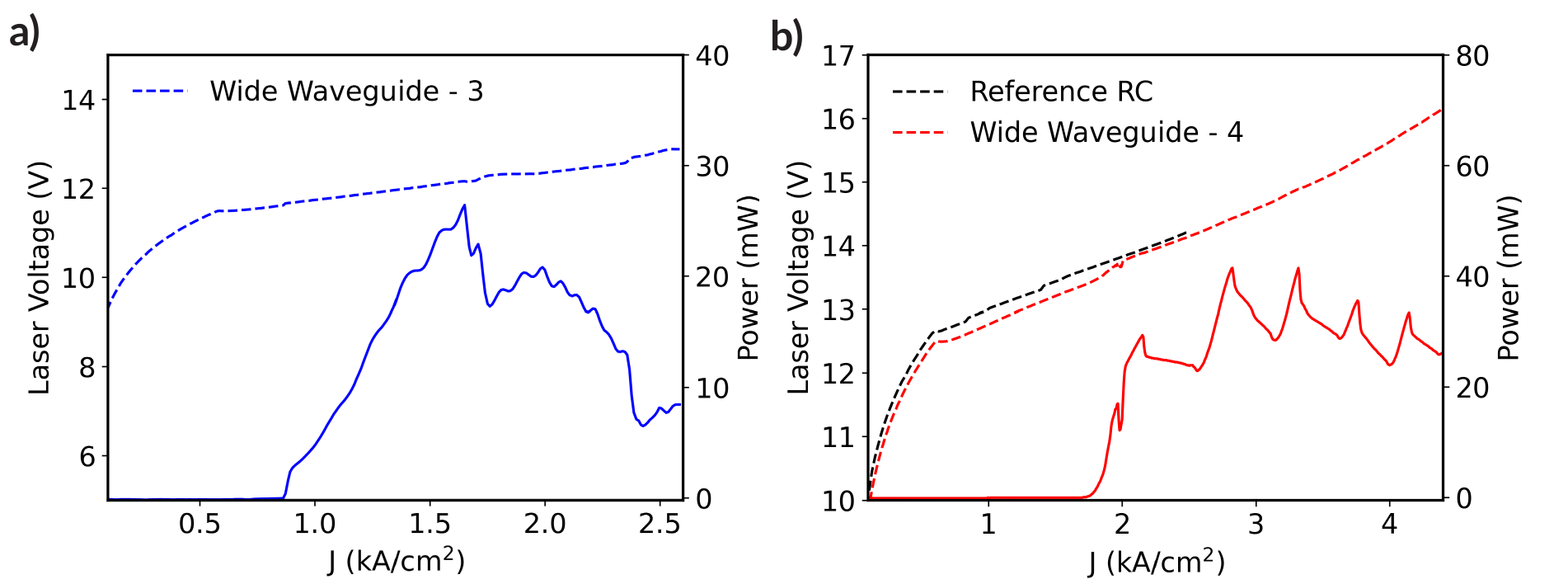}
  \caption{\textbf{b)} Light-Current density and Power-Current density of the wide waveguide device used to measure the spectra in Fig.\ref{fig:spectra}.b. of the main text. \textbf{a)} Comparison of Voltage-Current density curves between a reference racetrack (black) and a wide waveguide laser (red) processed on the same active material (EV3032). For the latter, the Power-Current density curve is also reported. }
  \label{SI:LIV}
\end{figure}

Every Light-Current-Power measurements is performed at a heat-sink temperature of 253 °K and with a DC current source. For the sake of completeness, we report in S.\ref{SI:LIV}.a the light-current-voltage characteristic of the wide waveguide-3 laser, which is the one used to measure the spectra present in Fig.\ref{fig:spectra}.b. The lasers wide waveguide-1-2-3 are all processed on the layer EV2547.

We also report in S.\ref{SI:LIV}.b, a direct comparison of the voltage-current curves between a reference racetrack and the wide waveguide-4 device, processed on the same layer (EV3032) to explicitly show that the introduction of an additional waveguide does not impact the current threshold, indicating that the MOVPE regrowth was optimal. We underline that, since the active region EV2547 is mostly based on the design of EV3032, their electrical behavior results similar, therefore the current threshold of all the wide waveguide devices are comparable. Instead, the narrow waveguide devices are processed using the active material EV1426 which strongly differs from the previous design, therefore, the higher current is due to the active material design and not to the top waveguide, which, as mentioned, does not introduce additional losses. In the following part, we report the details of the structures in \AA , where we indicate in normal text the InGaAs well and in bold the AlInAs barrier thickness, starting from the injection barrier.

\begin{itemize}
    \item \textbf{EV3032} The period is repeated 35 times. \newline
    Doping Sheet Density = 1.25e11 cm$^{-2}$ Composition: Ga = 32.6\%, Al = 65.2\%  \newline 
    \textbf{35}/11/\textbf{13}/38/\textbf{10}/35/\textbf{18}/27/\textbf{19}/26/\textbf{15}/23/\textbf{14}/21/\textbf{22}/19/\textbf{20}/19/\textbf{19}/17/\textbf{24}/17

    \item \textbf{EV1426} The period is repeated 35 times. \newline
    Doping Sheet Density = 1.1e11 cm$^{-2}$ Composition: Ga = 38\%, Al = 60\%  \newline 
    \textbf{39}/13/\textbf{14}/41/\textbf{17}/37/\textbf{25}/28/\textbf{15}/26/\textbf{16}/24/\textbf{17}/22/\textbf{19}/21/\textbf{21}/20/\textbf{23}/18/\textbf{24}/18

    \item \textbf{EV2547} The active region is a dual stack. AR1 is repeated 18 times and it is identical to EV3032 with a sheet density of 1e11cm$^{-2}$. AR2 is repeated 17 times and has the following structure:
    Doping Sheet Density = 0.992e11 cm$^{-2}$ Composition: Ga = 31.6\%, Al = 66.5\%  \newline 
    \textbf{35}/13.1/\textbf{14.8}/37.6/\textbf{10.3}/32.9/\textbf{19.9}/27.6/\textbf{14}/\textbf{24.1}/ \newline
    \textbf{14.8}/24.6/\textbf{13.3}/21.9/\textbf{15.8}/18.5/\textbf{19.7}/17.4/\textbf{20.6}/14.7/\textbf{21.7}/15.6
    
\end{itemize}

\begin{figure}[htb!]
  \centering
  \includegraphics[width=.9\textwidth]{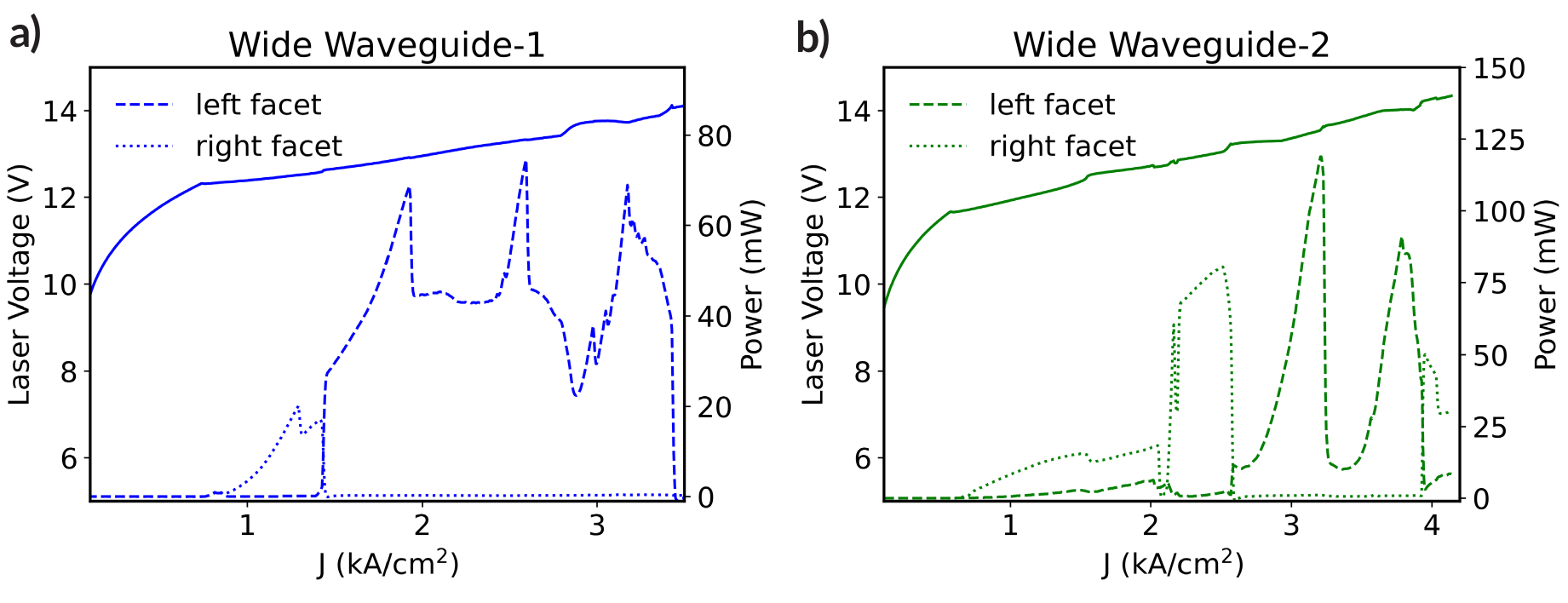}
  \caption{ Light-Current density - Power curves for the device wide waveguide-1 \textbf{a)} and wide waveguide-2 \textbf{b)}. for each curve we report the power collected from the left and right extraction facet of the passive waveguide.}
  \label{SI:LIV2}
\end{figure}

In S.\ref{SI:LIV2} we report the Light-Current-Voltage curves for the two wide waveguide devices already mentioned in the main text, adding the power extracted by both passive waveguide facets.
This further underlines the extremely high oscillating behavior of the power, indicating that there are current points where the power is not disappearing but the lasing direction in the ring is switching. This is the reason why, even if in the voltage current curve the lasing threshold is evident, it is not possible to observe it in the power curve.

\section*{Simulation Details}

\begin{figure}[htb!]
  \centering
  \includegraphics[width=.9\textwidth]{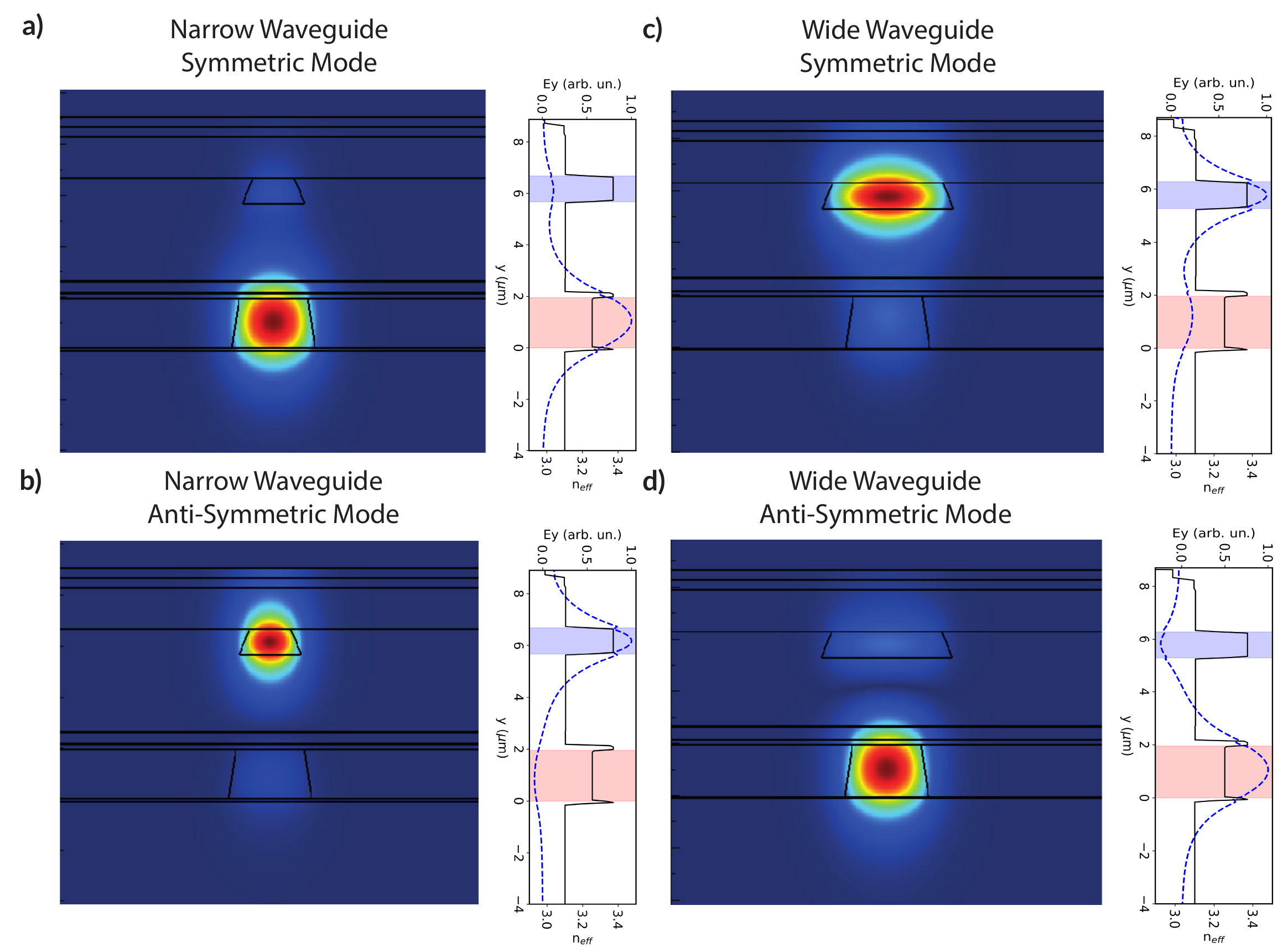}
  \caption{2D simulation of the coupled waveguides symmetric and anti-symmetric mode. For each of the two cases, the narrow \textbf{a-b)} and wide waveguide \textbf{c-d)} are reported. Each inset represents the mode profile along the y direction in the center of the cavity (x=0) together with the refractive index profile. }
  \label{SI:Sim1}
\end{figure}

We report in S.\ref{SI:Sim1} the simulated supermode shape for the symmetric and antisymmetric solution, for the two cases of narrow and wide waveguide. As mentioned in the main text, besides the inversion between the lasing mode which passes from the symmetric to the anti-symmetric changing from the narrow to wide waveguide, the mode shape results similar in the two cases. In particular, the mode in the wide waveguide laser does not present a higher order mode in the passive element, indicating that the higher order mode measured in the far field must come from the mode mismatch at the interface change between the coupled waveguide region and the purely passive region.

\begin{figure}[htb!]
  \centering
  \includegraphics[width=.9\textwidth]{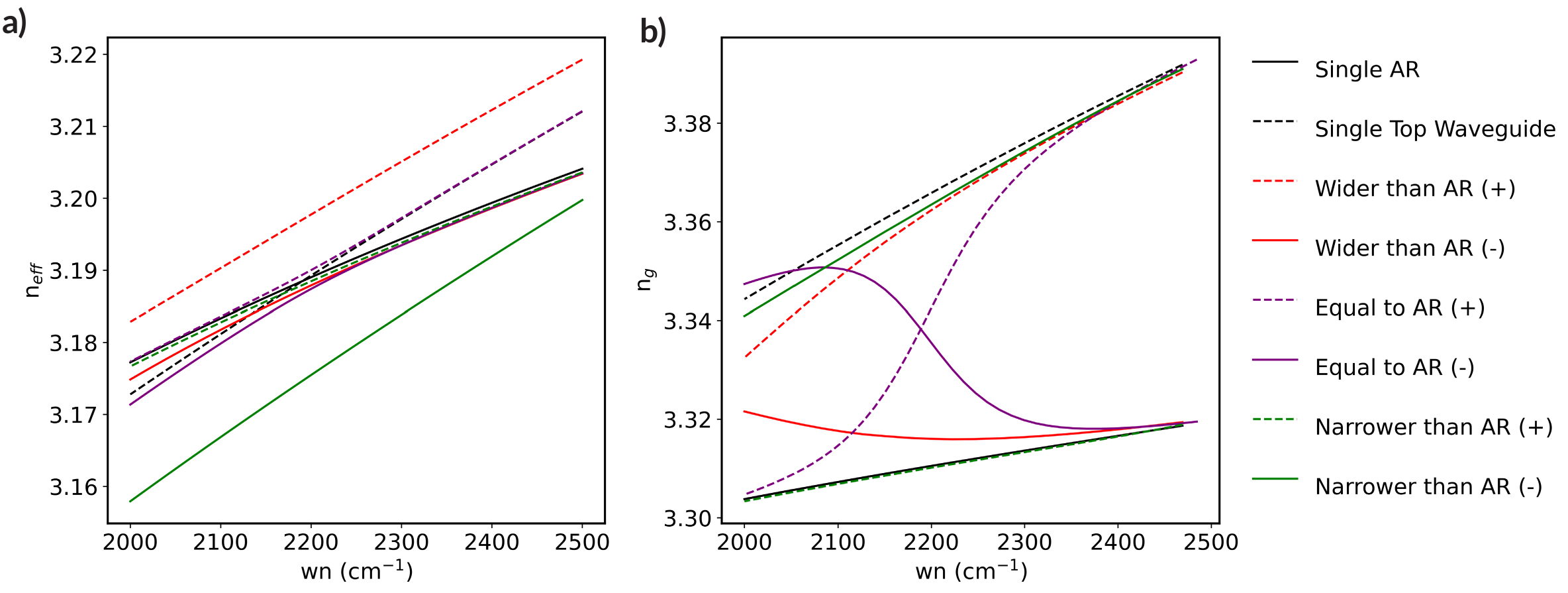}
  \caption{Effective refractive index \textbf{a)} and group index \textbf{b)} for the two uncoupled waveguides and for the cases of a wider, narrower and same size top waveguide compared to the active region width. Both symmetric (+) and antisymmetric (-) solutions are displayed. }
  \label{SI:Sim2}
\end{figure}

The dispersion simulations reported in Fig.\ref{fig:dispersion} of the main text are performed using the 2D solver MODE of Ansys-Lumerical, assuming a device structure like the one in Tab.\ref{tab:dw}, an active region width of 5 $\mu$m and a passive waveguide width of 3 $\mu$m for the narrow waveguide case and 8.5 $\mu$m for the wide waveguide case. The refractive indices of the different materials are chosen from \cite{d_palik_optical_1997}. The change in refractive index and losses due to the different doping are taken into account using the usual Drude formula. From the solver it is possible to obtain the effective refractive index n$_{eff}$ of TM polarized field for the symmetric and anti-symmetric mode, which we report in S.\ref{SI:Sim2}.a. From n$_{eff}$ we compute the group index n$_g$, reported in S.\ref{SI:Sim2}.b and the GVD reported in the main text.

\section*{Anti-reflection coating }

We report, as an example, the CW spectral map and the RF frequency sweep spectral map for the narrow waveguide-2 device, also reported in the main text in Fig.\ref{fig:spectra}.c before and after applying an anti-reflection (AR) coating. 

\begin{figure}[htb!]
  \centering
  \includegraphics[width=.9\textwidth]{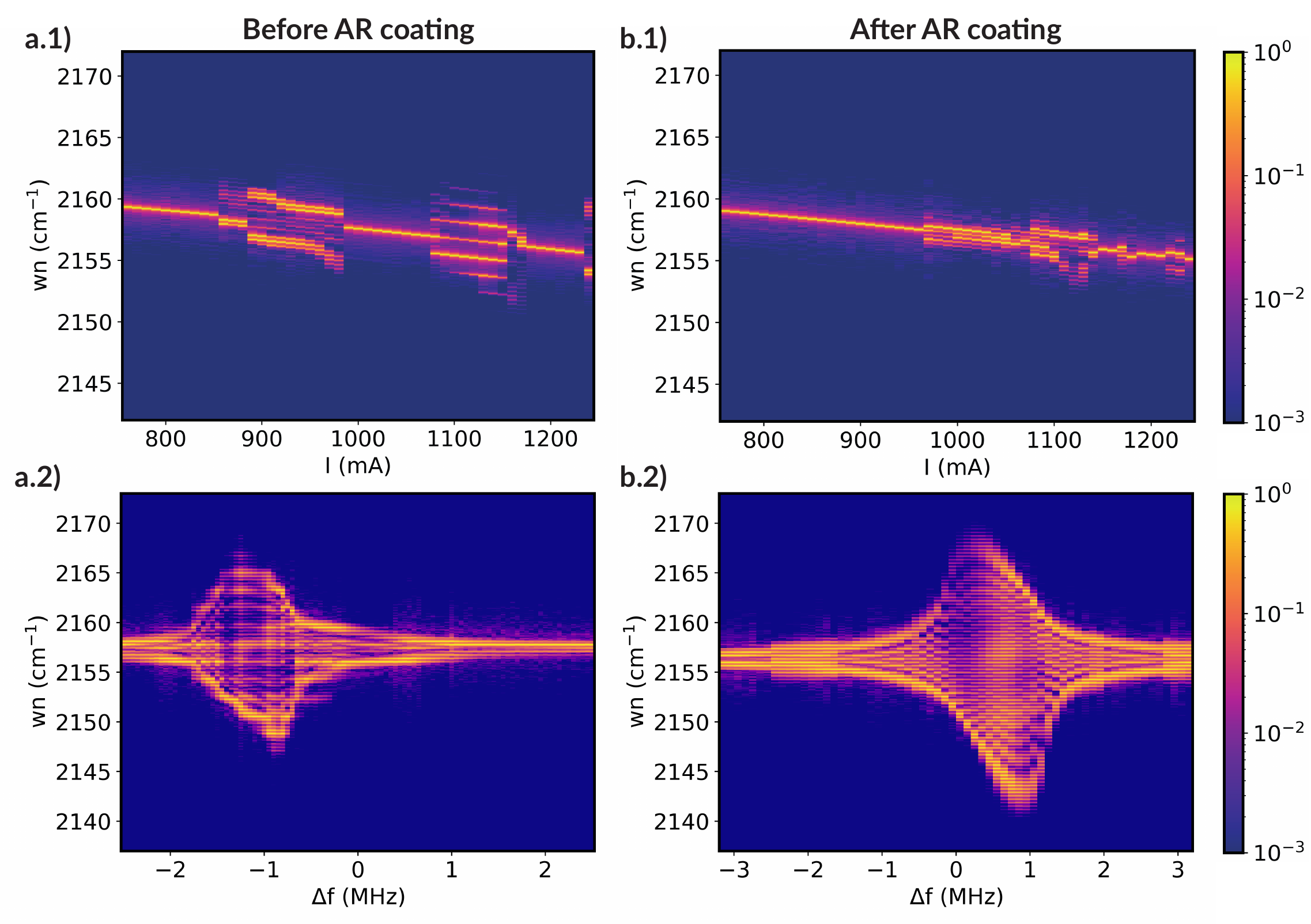}
  \caption{CW spectral map (first row) and RF frequency sweep spectral map (second row) for the NTW-1 device before \textbf{a)} and after \textbf{b)} AR coating.}
  \label{SI:spectra}
\end{figure}

The coating consists of a 700 nm thick Al$_2$O$_3$ film. As mentioned in the main text, the coating allows to reduce the facet reflectivity from 27 $\%$ to roughly 8$\%$,  and therefore the power reflected into the counterpropagating lasing direction. This is clear both from the CW spectral map, which presents less evident multimode states even if it is not possible to suppress them completely for the higher current range, and from the frequency map, that results in a cleaner and slightly broader spectrum. For the latter comparison, the DC bias point is the same (1.015 A for both cases), and the RF modulation power is chosen to be similar, in particular 23 dBm before coating and 24 dBm after coating. The repetition frequency in this case is $f_{rep} = 10.15$GHz.

\newpage

\bibliographystyle{ieeetr}
\bibliography{references}

\begin{thebibliography}{10}

\bibitem{fortier_20_2019}
T.~Fortier and E.~Baumann, ``20 years of developments in optical frequency comb technology and applications,'' {\em Communications Physics}, vol.~2, p.~153, Dec. 2019.

\bibitem{picque_frequency_2019}
N.~Picqué and T.~W. Hänsch, ``Frequency comb spectroscopy,'' {\em Nature Photonics}, vol.~13, pp.~146--157, Mar. 2019.

\bibitem{hussein_passive_2024}
H.~M.~E. Hussein, S.~Kim, M.~Rinaldi, A.~Alù, and C.~Cassella, ``Passive frequency comb generation at radiofrequency for ranging applications,'' {\em Nature Communications}, vol.~15, p.~2844, Apr. 2024.
\newblock Publisher: Nature Publishing Group.

\bibitem{okawachi_chip-scale_2023}
Y.~Okawachi, B.~Y. Kim, M.~Lipson, and A.~L. Gaeta, ``Chip-scale frequency combs for data communications in computing systems,'' {\em Optica}, vol.~10, pp.~977--995, Aug. 2023.
\newblock Publisher: Optica Publishing Group.

\bibitem{udem_optical_2002}
T.~Udem, R.~Holzwarth, and T.~W. Hänsch, ``Optical frequency metrology,'' {\em Nature}, vol.~416, pp.~233--237, Mar. 2002.
\newblock Publisher: Nature Publishing Group.

\bibitem{ye_optical_2003}
J.~Ye, H.~Schnatz, and L.~Hollberg, ``Optical frequency combs: from frequency metrology to optical phase control,'' {\em IEEE Journal of Selected Topics in Quantum Electronics}, vol.~9, pp.~1041--1058, July 2003.

\bibitem{faist_quantum_1994}
J.~Faist, F.~Capasso, D.~L. Sivco, C.~Sirtori, A.~L. Hutchinson, and A.~Y. Cho, ``Quantum {Cascade} {Laser},'' {\em Science}, vol.~264, no.~5158, pp.~553--556, 1994.
\newblock \_eprint: https://www.science.org/doi/pdf/10.1126/science.264.5158.553.

\bibitem{hugi_mid-infrared_2012}
A.~Hugi, G.~Villares, S.~Blaser, H.~C. Liu, and J.~Faist, ``Mid-infrared frequency comb based on a quantum cascade laser,'' {\em Nature}, vol.~492, pp.~229--233, Dec. 2012.

\bibitem{burghoff_terahertz_2014}
D.~Burghoff, T.-Y. Kao, N.~Han, C.~W.~I. Chan, X.~Cai, Y.~Yang, D.~J. Hayton, J.-R. Gao, J.~L. Reno, and Q.~Hu, ``Terahertz laser frequency combs,'' {\em Nature Photonics}, vol.~8, pp.~462--467, June 2014.

\bibitem{burghoff_unraveling_2020}
D.~Burghoff, ``Unraveling the origin of frequency modulated combs using active cavity mean-field theory,'' {\em Optica}, vol.~7, p.~1781, Dec. 2020.

\bibitem{opacak_theory_2019}
N.~Opačak and B.~Schwarz, ``Theory of {Frequency}-{Modulated} {Combs} in {Lasers} with {Spatial} {Hole} {Burning}, {Dispersion}, and {Kerr} {Nonlinearity},'' {\em Physical Review Letters}, vol.~123, p.~243902, Dec. 2019.

\bibitem{villares_dispersion_2016}
G.~Villares, S.~Riedi, J.~Wolf, D.~Kazakov, M.~J. Süess, P.~Jouy, M.~Beck, and J.~Faist, ``Dispersion engineering of quantum cascade laser frequency combs,'' {\em Optica}, vol.~3, p.~252, Mar. 2016.

\bibitem{kapsalidis_mid-infrared_2021}
F.~Kapsalidis, B.~Schneider, M.~Singleton, M.~Bertrand, E.~Gini, M.~Beck, and J.~Faist, ``Mid-infrared quantum cascade laser frequency combs with a microstrip-like line waveguide geometry,'' {\em Applied Physics Letters}, vol.~118, p.~071101, Feb. 2021.

\bibitem{meng_mid-infrared_2020}
B.~Meng, M.~Singleton, M.~Shahmohammadi, F.~Kapsalidis, R.~Wang, M.~Beck, and J.~Faist, ``Mid-infrared frequency comb from a ring quantum cascade laser,'' {\em Optica}, vol.~7, p.~162, Feb. 2020.

\bibitem{jaidl_comb_2021}
M.~Jaidl, N.~Opačak, M.~A. Kainz, S.~Schönhuber, D.~Theiner, B.~Limbacher, M.~Beiser, M.~Giparakis, A.~M. Andrews, G.~Strasser, B.~Schwarz, J.~Darmo, and K.~Unterrainer, ``Comb operation in terahertz quantum cascade ring lasers,'' {\em Optica}, vol.~8, p.~780, June 2021.

\bibitem{lugiato_nonlinear_2015}
L.~Lugiato, {\em Nonlinear {Optical} {Systems}}.
\newblock West Nyack: Cambridge University Press, 1st ed~ed., 2015.

\bibitem{meng_dissipative_2022}
B.~Meng, M.~Singleton, J.~Hillbrand, M.~Franckié, M.~Beck, and J.~Faist, ``Dissipative {Kerr} solitons in semiconductor ring lasers,'' {\em Nature Photonics}, vol.~16, pp.~142--147, Feb. 2022.
\newblock Number: 2 Publisher: Nature Publishing Group.

\bibitem{kazakov_active_2024}
D.~Kazakov, T.~P. Letsou, M.~Beiser, Y.~Zhi, N.~Opačak, M.~Piccardo, B.~Schwarz, and F.~Capasso, ``Active mid-infrared ring resonators,'' {\em Nature Communications}, vol.~15, p.~607, Jan. 2024.
\newblock Publisher: Nature Publishing Group.

\bibitem{opacak_nozakibekki_2024}
N.~Opačak, D.~Kazakov, L.~L. Columbo, M.~Beiser, T.~P. Letsou, F.~Pilat, M.~Brambilla, F.~Prati, M.~Piccardo, F.~Capasso, and B.~Schwarz, ``Nozaki–{Bekki} solitons in semiconductor lasers,'' {\em Nature}, vol.~625, pp.~685--690, Jan. 2024.
\newblock Publisher: Nature Publishing Group.

\bibitem{micheletti_terahertz_2023}
P.~Micheletti, U.~Senica, A.~Forrer, S.~Cibella, G.~Torrioli, M.~Frankié, M.~Beck, J.~Faist, and G.~Scalari, ``Terahertz optical solitons from dispersion-compensated antenna-coupled planarized ring quantum cascade lasers,'' {\em Science Advances}, vol.~9, p.~eadf9426, June 2023.
\newblock Publisher: American Association for the Advancement of Science.

\bibitem{seitner_backscattering-induced_2024}
L.~Seitner, J.~Popp, I.~Heckelmann, R.-E. Vass, B.~Meng, M.~Haider, J.~Faist, and C.~Jirauschek, ``Backscattering-{Induced} {Dissipative} {Solitons} in {Ring} {Quantum} {Cascade} {Lasers},'' {\em Physical Review Letters}, vol.~132, p.~043805, Jan. 2024.

\bibitem{heckelmann_quantum_2023}
I.~Heckelmann, M.~Bertrand, A.~Dikopoltsev, M.~Beck, G.~Scalari, and J.~Faist, ``Quantum walk comb in a fast gain laser,'' {\em Science}, vol.~382, pp.~434--438, Oct. 2023.
\newblock Publisher: American Association for the Advancement of Science.

\bibitem{dikopoltsev_theory_nodate}
A.~Dikopoltsev, I.~Heckelmann, B.~Schneider, M.~Bertrand, and J.~Faist, ``The theory of the quantum walk comb laser,'' {\em Nanophtonics}, vol.~in press.

\bibitem{letsou_high-power_2025}
T.~P. Letsou, J.~Fuchsberger, N.~Opačak, D.~Kazakov, B.~Schwarz, and F.~Capasso, ``High-power quantum walk frequency combs,'' Feb. 2025.
\newblock arXiv:2502.10919 [physics].

\bibitem{lin_mid-infrared_2018}
H.~Lin, Z.~Luo, T.~Gu, L.~C. Kimerling, K.~Wada, A.~Agarwal, and J.~Hu, ``Mid-infrared integrated photonics on silicon: a perspective,'' {\em Nanophotonics}, vol.~7, pp.~393--420, Feb. 2018.
\newblock Publisher: De Gruyter.

\bibitem{spott_quantum_2016}
A.~Spott, J.~Peters, M.~L. Davenport, E.~J. Stanton, C.~D. Merritt, W.~W. Bewley, I.~Vurgaftman, C.~S. Kim, J.~R. Meyer, J.~Kirch, L.~J. Mawst, D.~Botez, and J.~E. Bowers, ``Quantum cascade laser on silicon,'' {\em Optica}, vol.~3, pp.~545--551, May 2016.
\newblock Publisher: Optica Publishing Group.

\bibitem{pelucchi_potential_2022}
E.~Pelucchi, G.~Fagas, I.~Aharonovich, D.~Englund, E.~Figueroa, Q.~Gong, H.~Hannes, J.~Liu, C.-Y. Lu, N.~Matsuda, J.-W. Pan, F.~Schreck, F.~Sciarrino, C.~Silberhorn, J.~Wang, and K.~D. Jöns, ``The potential and global outlook of integrated photonics for quantum technologies,'' {\em Nature Reviews Physics}, vol.~4, pp.~194--208, Mar. 2022.
\newblock Publisher: Nature Publishing Group.

\bibitem{hinkov_mid-infrared_2022}
B.~Hinkov, F.~Pilat, L.~Lux, P.~L. Souza, M.~David, A.~Schwaighofer, D.~Ristanić, B.~Schwarz, H.~Detz, A.~M. Andrews, B.~Lendl, and G.~Strasser, ``A mid-infrared lab-on-a-chip for dynamic reaction monitoring,'' {\em Nature Communications}, vol.~13, p.~4753, Aug. 2022.
\newblock Publisher: Nature Publishing Group.

\bibitem{wang_monolithic_2022}
R.~Wang, P.~Täschler, Z.~Wang, E.~Gini, M.~Beck, and J.~Faist, ``Monolithic {Integration} of {Mid}-{Infrared} {Quantum} {Cascade} {Lasers} and {Frequency} {Combs} with {Passive} {Waveguides},'' {\em ACS Photonics}, vol.~9, pp.~426--431, Feb. 2022.
\newblock Publisher: American Chemical Society.

\bibitem{jung_homogeneous_2019}
S.~Jung, D.~Palaferri, K.~Zhang, F.~Xie, Y.~Okuno, C.~Pinzone, K.~Lascola, and M.~A. Belkin, ``Homogeneous photonic integration of mid-infrared quantum cascade lasers with low-loss passive waveguides on an {InP} platform,'' {\em Optica}, vol.~6, pp.~1023--1030, Aug. 2019.
\newblock Publisher: Optica Publishing Group.

\bibitem{burghart_multi-color_2025}
D.~Burghart, K.~Zhang, W.~Oberhausen, A.~Köninger, G.~Boehm, and M.~A. Belkin, ``Multi-color photonic integrated circuits based on homogeneous integration of quantum cascade lasers,'' {\em Nature Communications}, vol.~16, p.~3563, Apr. 2025.
\newblock Publisher: Nature Publishing Group.

\bibitem{montesinos-ballester_low-loss_2024}
M.~Montesinos-Ballester, E.~Jöchl, V.~Turpaud, J.~Hillbrand, M.~Bertrand, D.~Marris-Morini, E.~Gini, and J.~Faist, ``Low-{Loss} {Buried} {InGaAs}/{InP} {Integrated} {Waveguides} in the {Long}-{Wave} {Infrared},'' {\em ACS Photonics}, vol.~11, pp.~2236--2241, June 2024.
\newblock Publisher: American Chemical Society.

\bibitem{bidaux_coupled-waveguides_2018}
Y.~Bidaux, F.~Kapsalidis, P.~Jouy, M.~Beck, and J.~Faist, ``Coupled-{Waveguides} for {Dispersion} {Compensation} in {Semiconductor} {Lasers},'' {\em Laser \& Photonics Reviews}, vol.~12, no.~5, p.~1700323, 2018.
\newblock \_eprint: https://onlinelibrary.wiley.com/doi/pdf/10.1002/lpor.201700323.

\bibitem{beck_continuous_2002}
M.~Beck, D.~Hofstetter, T.~Aellen, J.~Faist, U.~Oesterle, M.~Ilegems, E.~Gini, and H.~Melchior, ``Continuous {Wave} {Operation} of a {Mid}-{Infrared} {Semiconductor} {Laser} at {Room} {Temperature},'' {\em Science}, vol.~295, pp.~301--305, Jan. 2002.
\newblock Publisher: American Association for the Advancement of Science.

\bibitem{opacak_impact_2024}
N.~Opačak, B.~Schneider, J.~Faist, and B.~Schwarz, ``Impact of higher-order dispersion on frequency-modulated combs,'' {\em Optics Letters}, vol.~49, pp.~794--797, Feb. 2024.
\newblock Publisher: Optica Publishing Group.

\bibitem{elshaari_hybrid_2020}
A.~W. Elshaari, W.~Pernice, K.~Srinivasan, O.~Benson, and V.~Zwiller, ``Hybrid integrated quantum photonic circuits,'' {\em Nature Photonics}, vol.~14, pp.~285--298, May 2020.
\newblock Publisher: Nature Publishing Group.

\bibitem{taschler_short_2023}
P.~Täschler, L.~Miller, F.~Kapsalidis, M.~Beck, and J.~Faist, ``Short pulses from a gain-switched quantum cascade laser,'' {\em Optica}, vol.~10, pp.~507--512, Apr. 2023.
\newblock Publisher: Optica Publishing Group.

\bibitem{d_palik_optical_1997}
E.~D.~Palik, ``Optical {Parameters} for the {Materials} in \textit{{HOC} {I}, {HOC} {II}}, and \textit{{HOC} {III}},'' in {\em Handbook of {Optical} {Constants} of {Solids}} (E.~D. Palik, ed.), pp.~187--227, Burlington: Academic Press, Jan. 1997.

\end{thebibliography}

\end{document}